\title{The SiTian Project}
\author{
        JiFeng Liu \\
        Key Laboratory of Optical Astronomy\\National Astronomical Observatories\\Chinese Academy of Sciences\\Beijing, \underline{China}\\{\tt jfliu@nao.cas.cn}
            \and
        Roberto Soria\\
        School of Astronomy and Space Sciences\\
        University of the Chinese Academy of Sciences\\
        Beijing, \underline{China}\\{\tt rsoria@nao.cas.cn}
        \and
        Xue-Feng Wu \\
        Purple Mountain Observatory \\ Chinese Academy of Sciences\\Nanjing, \underline{China}
        \and 
        Hong Wu \\
        National Astronomical Observatories\\Chinese Academy of Sciences\\Beijing, \underline{China}
        \and
        Zhaohui Shang \\
        National Astronomical Observatories\\Chinese Academy of Sciences\\Beijing, \underline{China}
        \and
        on behalf of the SiTian team
}
\date{\today}

\documentclass[12pt]{article}
\usepackage{graphicx}
\usepackage{multicol}
\usepackage{multirow}
\usepackage{rotating}

\begin{document}
\maketitle

\textbf{Accepted by the Journal ``Annals of the Brazilian Academy of Sciences'' as part of the Proceedings for the BRICS Astronomy Workshop  - BAWG 2019.}

\begin{abstract}
SiTian is an ambitious ground-based all-sky optical monitoring project, developed by the Chinese Academy of Sciences. The concept is an integrated network of dozens of 1-m-class telescopes deployed partly in China and partly at various other sites around the world. The main science goals are the detection, identification and monitoring of optical transients (such as gravitational wave events, fast radio bursts, supernovae) on the largely unknown timescales of less than 1 day; SiTian will also provide a treasure trove of data for studies of AGN, quasars, variable stars, planets, asteroids, and microlensing events. To achieve those goals, SiTian will scan at least 10,000 square deg of sky every 30 min, down to a detection limit of $V \approx 21$ mag. The scans will produce simultaneous light-curves in 3 optical bands. In addition, SiTian will include at least three 4-m telescopes specifically allocated for follow-up spectroscopy of the most interesting targets. We plan to complete the installation of 72 telescopes by 2030 and start full scientific operations in 2032.
\end{abstract}

\section{Introduction}

All-sky optical monitoring of transient and variable sources is one of the most promising lines of research over the next decade. Rapid detection and follow-up studies of optical counterparts is crucial for the precise localization and physical interpretation of various classes of high energy transients, such as neutron star mergers, tidal disruption events, fast radio bursts.  

There are already many time-domain sky surveys in operation or construction. Some are based on a single telescope with an extremely wide field of view. For example, the Zwicky Transient Factory (ZTF) at Palomar Observatory uses a 1.2-m Schmidt telescope with a 47 square deg field of view \cite{bellm19,graham19}.
The Rubin Observatory's Legacy Survey of Space and Time (LSST) will use an 8-m class telescope (effective collecting area of a 6.7-m filled aperture) with a 9.6 square deg field of view \cite{ivezic19}. From 2023, it will survey the southern sky from La Serena (Chile), down to $g \approx 25$ mag per scan. However, it will only have a few day cadence, not suitable for the study of fast transients on timescales of about 1 hour to 1 day.
Other monitoring projects are based on networks of telescopes. For example, the Las Cumbres Observatory (LCO) consists of 23 robotic telescopes (with a mix of 2-m, 1-m and 40-cm apertures) with imaging and spectroscopic capabilities, installed in several northern and southern hemisphere sites \cite{brown13}; one limitation of the LCO network is the relatively small field of view (only $26^{\prime} \times 26^{\prime}$ on the 1-m telescopes). The All-Sky Automated Survey for Supernovae (ASAS-SN) deploys 6 units of 4 telescopes each, also located in several continents, with a full coverage of the whole sky every night \cite{kochanek17}; the main limitations of the ASAS-SN survey are the small telescope size (only 14-cm apertures, for a detection limit of $V \approx 18$ mag) and the large pixel size ($7^{\prime\prime}.8$), not suitable for crowded fields and for accurate astrometry.

Based on the experience and limitations of these and other surveys, the Chinese astronomical community agreed to invest money and resources for an ambitious and innovative sky monitoring project, based on an integrated, world-wide network of telescopes and on a fast (sub-hour) cadence; the project is provisionally called SiTian (``Sky Monitoring"). The international nature of this project and its benefits outside the specific research field ({\it e.g.}, in terms of science outreach and big data management) are well suited to the research priorities of the BRICS countries. Indeed, the ambitious proposed BRICS Flagship project on transients (see these Proceedings) will include the proposed Sitian network in its future development.

\section{Structure of the SiTian project}
The project will consists of four components (Fig.~1): 

\begin{itemize}
\item SiTian Brain: an intelligent integrated control system. Its tasks are: a) scheduling and coordination of the observations of the various telescopes, so that they can most effectively divide their time between regular scans and targets of opportunity; b) real-time processing and preliminary analysis of a massive imaging dataset (expected to reach 100 PB per year), including real-time identification of follow-up targets, immediate communication of transient alerts, and data storage in easily accessible archive formats.
\item SiTian Array, composed of a number of ``units", each of which made of three 1-m-class Schmidt telescopes equipped with scientific Complementary Metal-Oxide-Semiconductor (sCMOS) detectors. The three telescopes of each unit will simultaneously observe the same target, in three different filters ($u$, $g$, $i$); a small number of telescopes may be equipped with an H$\alpha$ filter. The field of view of each telescope is $5 \times 5$ square deg. Each field will be observed for 1 min, and the unit will then move on to an adjacent field, in a pre-planned sequence.
\item Three or four 4-m-class telescopes for spectral identification and follow-up studies. Within a few minutes, we will obtain low/medium resolution spectra of transient sources identified with the imaging telescopes, down to 20 mag. We envisage that the imagers will have a field of view of $5^{\prime} \times 5^{\prime}$ with an image-plate scale of $0^{\prime \prime}.3$/pixel; the long-slit spectrographs will offer resolutions between $\approx$1000 and 5000. At least one of these telescopes will be equipped with a large-field-of-view, multi-fiber spectrograph similar to the one built for LAMOST (at the Xinglong Observatory near Beijing). Possible sites for the 4-m-class telescopes are Muztagh-Ata (Xinjiang Province, on the western side of the Tibetan plateau), Ali (central Tibet) and Lenghu (Qinghai Province).
\item Use of allocated or target-of-opportunity time on existing telescopes (mostly in the 1-m and 2-m class) and planned future telescopes for more extended follow-up studies of particular sources. In particular, we envisage that SiTian will become operational around the same time as the 12-meter Large Optical-Infrared Telescope (LOT), which is planned to be located in Muztagh-Ata.

\end{itemize}

\vspace{0.1cm}
\noindent
{\large{{\bf{Technical Specifications}}}}\\
\noindent
The number of SiTian units is the key factor that determines the depth and cadence of the survey. The goal is to scan 10,000 square deg of sky every 30 min, down to a 5-$\sigma$ detection limit of $V \approx 21$ mag, with 1\% photometric accuracy at $V \approx 16$ mag, simultaneously in three colors. The requirement of simultaneous colours is the reason why each unit contains three telescopes (one per filter) pointing at the same area of sky. The alternative solution of using a single telescope with dichroics feeding three cameras is not suitable, because it introduces image distortions in the large field of view of our telescopes, and also more optical complexity and cost. A single 1-m-class telescope can reach a detection limit of 21 mag in 1 min, and can cover $30 \times 25 = 750$ square deg in 30 min (neglecting overheads). To cover the target field of 10,000 square deg in three colours, we need at least 14 units (42 telescopes). A more ambitious plan we are currently developing is to extend our survey to a large fraction of both the northern and the southern sky. For this goal, additional units must be located in the southern hemisphere. One option we are looking at is to do a high-cadence scan (30-min or 45-min cadence) over 10,000 square deg, and a lower-cadence scan over 30,000 square deg, depending on how many units we can install in southern hemisphere sites. We envisage that the world-wide network will eventually include 24 units (72 telescopes).

In addition to the data from every scan, we will obtain deeper images of the fields by coadding the data from several days and weeks. For example, a co-added image including one week of data in all filters will reach about 23.5 mag, while for a one-year stack, we can reach 25 mag. In practice, the detection limit for stacked data will depend on the seeing conditions, on how many nights are lost to bad weather, and on the confusion limit for crowded fields.

The Nanjing Institute of Astronomical Optics and Technology is responsible for the overall technical design of the SiTian telescopes. That institute has already successfully developed major national projects such as LAMOST and the robotic Antarctic Survey Telescopes; they have vast experience in optical design, mirror grinding and polishing, high-precision driving and control. The National Astronomical Observatory's Laboratory of Optical and Infrared Detectors is in charge of developing and testing the sCMOS detectors.

\begin{figure}
\begin{center}
\includegraphics[width=11.5cm]{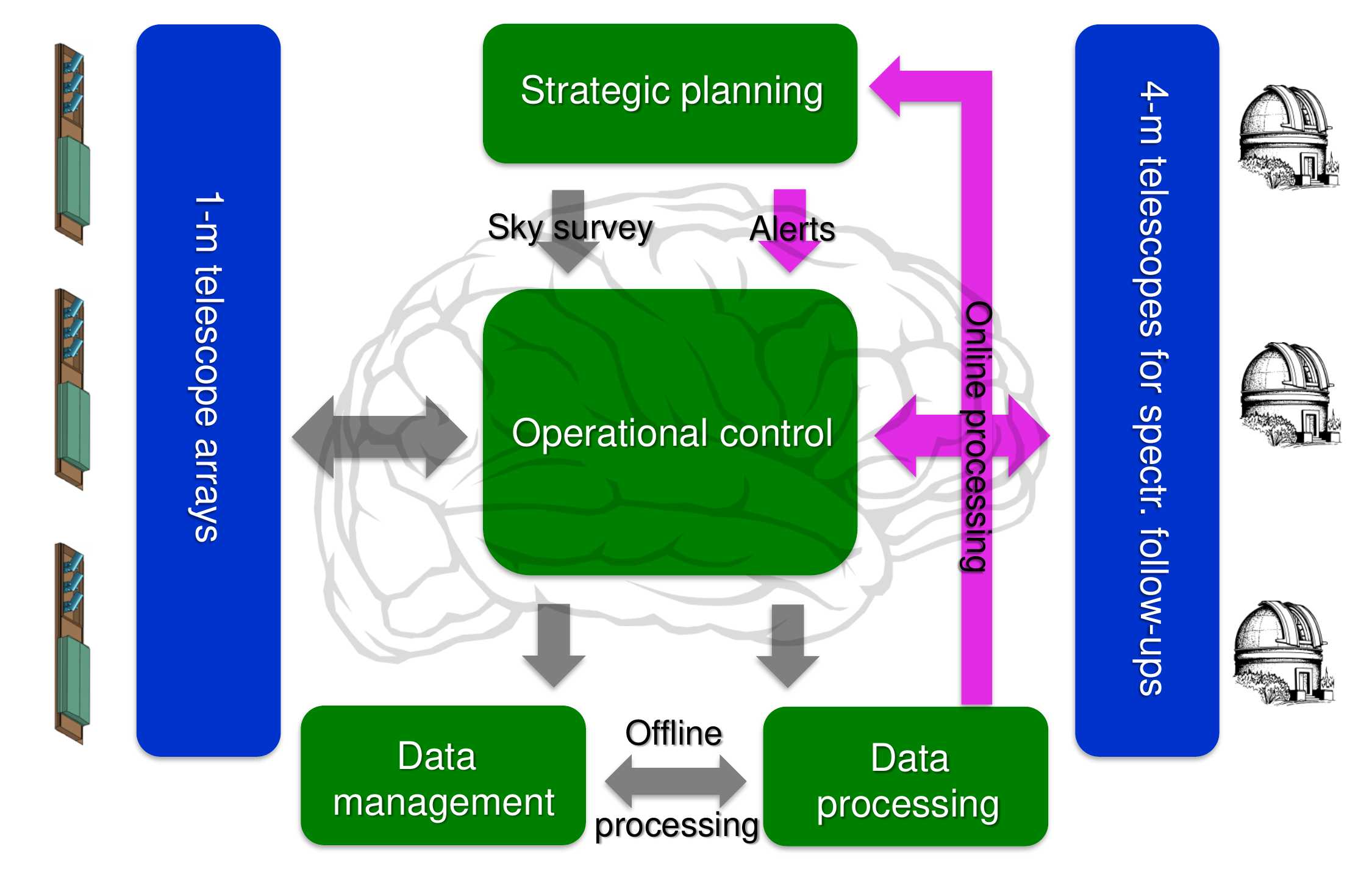}
\caption{Structure of the SiTian project, with the connections between its main elements: the monitoring array, the spectroscopic follow-up telescopes, and the data processing/management brain.}
\end{center}
\vspace{-0.5cm}
\end{figure}

\begin{sidewaystable}
  \begin{center}
  \footnotesize{
    \caption{Comparison between SiTian and a few other wide-field monitoring facilities}
    \vspace{0.3cm}
    \begin{tabular}{|l|c|c|c|c|c|c|c|c|r|}
    \hline
    & & & \multicolumn{2}{c|}{} & & & & & \\
      \multirow{2}{*}{\textbf{Facility}} & \multirow{2}{*}{\textbf{Aperture}} & \textbf{Number of} & \multicolumn{2}{c|}{\textbf{Detection limit}}     &  \textbf{FOV Single}  &  \textbf{Sky}  &  \multirow{2}{*}{\textbf{Cadence}}  & \multirow{2}{*}{\textbf{Site}} &  \multirow{2}{*}{\textbf{Year}}\\
       &  & \textbf{Telescopes} 
       & \textbf{Scan}  &  \textbf{Stacked}  &  \textbf{Exposure}  & \textbf{Coverage}   &    &    &\\
       & \textbf{(cm)} &  
       & \textbf{(mag)}  &  \textbf{(mag)}  &  \textbf{(deg$^2$)}  &  \textbf{(deg$^2$)}   &    &    &\\[5pt]
      \hline
          & & & & & & & & & \\
     \multirow{2}{*}{\textbf{SiTian}} & {\textbf{100}} & {\textbf{72}} & {\textbf{21.1}}  &  {\textbf{26.1}}\footnote{See Table 2 for details}  & {\textbf{600}}   &  {\textbf{10,000}}\footnote{An additional 20,000 square deg will be surveyed at slightly lower cadence (a few hr), with the SiTian units outside China.}  & {\textbf{30 min}}   &  China  & \multirow{2}{*}{\textbf{2032--}} \\
       & {\textbf{400}} & {\textbf{3}} & {\textbf{26.0}}  &  --  &  --  &  --  &  --  &  + global  &  \\[5pt]
       \hline
          & & & & & & & & & \\
       LSST & 670 & 1 & 24.8  & 27.5   &  1.6  &  20,000  &  1 wk  &  Chile  & 2022-- \\[5pt]
       \hline
          & & & & & & & & & \\
       Mephisto\footnote{http://www.swifar.ynu.edu.cn/info/1015/1073.htm} & 160 & 1 & 22.7  & 24.7   &  3.14  & 26,000  & $\sim$10 wk   &  China  &  2022--\\[5pt]
       \hline
          & & & & & & & & & \\
       Pan-STARRS\footnote{Magnitude limits based on Pan-STARRS1 DR2, not including the second telescope yet.} & 180 & 2 & 22.0  & 23.3   &  7.0  &  30,000  &  1 wk  &  Hawaii  & 2010-- \\[5pt]
       \hline
          & & & & & & & & & \\
       Sky Mapper\footnote{http://skymapper.anu.edu.au/} & 135 & 1 & 17.7  &  22.6  &  5.2  &  20,000  & 1 wk   &  Australia  & 2014-- \\[5pt]
       \hline
          & & & & & & & & & \\
       WFST\footnote{http://www.icehap.chiba-u.jp/amon2019/slides/15.XianZhong\_ZHENG.pdf} & 250 & 1 & 22.9  & 25.2   &  6.55  & 20,000   &  3 d  & China   & 2022-- \\[5pt]
       \hline
          & & & & & & & & & \\
       ZTF & 122 & 1 &  21.1 & 24.3   & 47.0   & 30,000   &  2 d  &  California  & 2017-- \\[5pt]
       \hline
    \end{tabular}
    \label{tab:table1}
    }
  \end{center}
\end{sidewaystable}

\vspace{0.3cm}
\noindent
{\large{{\bf{Data Management and Processing}}}}\\
\noindent
SiTian's hardware system is designed as a distributed system with on-site as well as remote operation and control centres, interconnected through a wide area network. The control software is a highly automated system, designed to coordinate the operation of each unit, to maximize the scanning efficiency and the observation time.
Data management and archiving will be based on a distributed storage system, which can be expanded horizontally; each unit will have a data service interface. Data processing will be divided into online and offline modes. Online (real-time) data processing will provide fast preliminary data analysis, sufficient for rapid source identification and classification, and for sending possible alerts, within one minute. This will be achieved thanks to parallel processing technology and high-speed multi-core computers, with deep learning algorithms ({\it e.g.}, neural networks and pattern recognition). Offline (non-real-time) data processing will provide high-precision measurements, more accurate images and light-curves, frame stacking, multiband comparisons, source catalogs, etc.

The SiTian telescopes will directly generate about 40 GB of data per minute, that is 24 TB every night; the processed data will be about six times the raw data, for an estimated total of about 140 TB per night. The data volume in one year will be around 40 PB. Taking into account redundancies and backups, we will need a storage capacity of at least 100 PB per year.

\begin{table}
  \begin{center}
  \footnotesize{
    \caption{Estimated limiting magnitudes in the stacked SiTian images}
    \vspace{0.3cm}
    \begin{tabular}{|l|c|c|}
    \hline
    & &  \\[-5pt]
{\textbf{Number of Scans}} & {\textbf{Total Exp Time}} &
{\textbf{{\it g}-band 5$\sigma$ Limit}}\\
   & {\textbf{(s)}} & {\textbf{(mag)}} \\[3pt]
         \hline
          & & \\[-5pt]
1 & 60 & 21.10\\[3pt]
       \hline
          & &  \\[-5pt]
3 & 180 & 21.70\\[3pt]
       \hline
          & & \\[-5pt]
10 (one night) & 600 & 22.35\\[3pt]
       \hline
          & & \\[-5pt]
300 (one month) & 18,000 & 24.20\\[3pt]
       \hline
          & & \\[-5pt]
1000 (one yr) & 60,000 & 24.85\\[3pt]
       \hline
          & & \\[-5pt]
10,000 (10 yr) & 600,000 & 26.10\\[3pt]
       \hline
    \end{tabular}
    \label{tab:table2}
    }
  \end{center}
\end{table}

\vspace{0.3cm}
\noindent
{\large{{\bf{Potential Locations}}}}\\
\noindent
For the units located in the Chinese territories (Fig.~2), we plan to install the first ones at a new observatory on the Saishiteng mountains, near the town of Lenghu, Qinghai Province (central China), at an elevation of about 4200 m, with excellent seeing. The official opening of the new observatory is scheduled for the second half of 2020. As an aside, Lenghu will also be the site of the 2.5-m Wide Field Survey Telescope (WFST), built by the University of Science and Technology of China, and expected to start operations in 2022; WFST will survey 20,000 square deg of the northern sky in $u$, $g$, $r$, $i$ and $z$, reaching 23$^{\rm{rd}}$ magnitude (and two magnitudes deeper in the 6-yr stacked image) for high-precision photometry and astrometry \cite{lou16}. Other SiTian array units will be installed at Muztagh-Ata and Ali. We are engaged in collaborative discussions to choose suitable sites in other northern hemisphere regions (Fig.~2), {\it e.g.}, at La Palma and at San Pedro Martir Observatory in Baja California (Mexico). 

For the southern hemisphere units (Fig.~2), a suitable site in Chile is Cerro Ventarrones (elevation 2890 m), about 30 km north-east of Cerro Paranal (site of the VLT). The Ventarrones site is available to SiTian through the South America Center for Astronomy of the Chinese Academy of Sciences (CASSACA). Moreover, we are optimistic about the possibility of locating a SiTian unit at Siding Spring Observatory (Australia), a site that is already hosting (or will soon host) a raft of other sky monitoring telescopes (the Australian National University's Sky Mapper \cite{keller07} and DREAMS \cite{soon20} telescopes, three LCO telescopes, a unit of the Gravitational-wave Optical Transient Observer). 

The third focal point of SiTian's southern hemisphere plan is South Africa, in collaboration with the South African Astronomical Observatory (SAAO). SAAO has begun (in 2020) a parallel project, aimed at networking its existing telescopes into an automated African Intelligent Observatory, or AIO, led by Dr Stephen Potter. This will form part of an eventual BRICS-wide network of existing and future telescopes, named the {\bf {BRICS Intelligent Telescope and Data Network}} (Lead Investigator: Prof. David Buckley), with one major aim being the detection and follow-up of transients. This proposed astronomy flagship project is an amalgamation of the BRICS Transients and Big Data proposals, both presented at this meeting (by Profs. David Buckley \& Russ Taylor, respectively).  SiTian will form the future major component of this network, allowing unprecedented cadence monitoring of the entire sky from facilities located in the BRICS countries.

\begin{figure}
\begin{center}
\includegraphics[width=13.5cm]{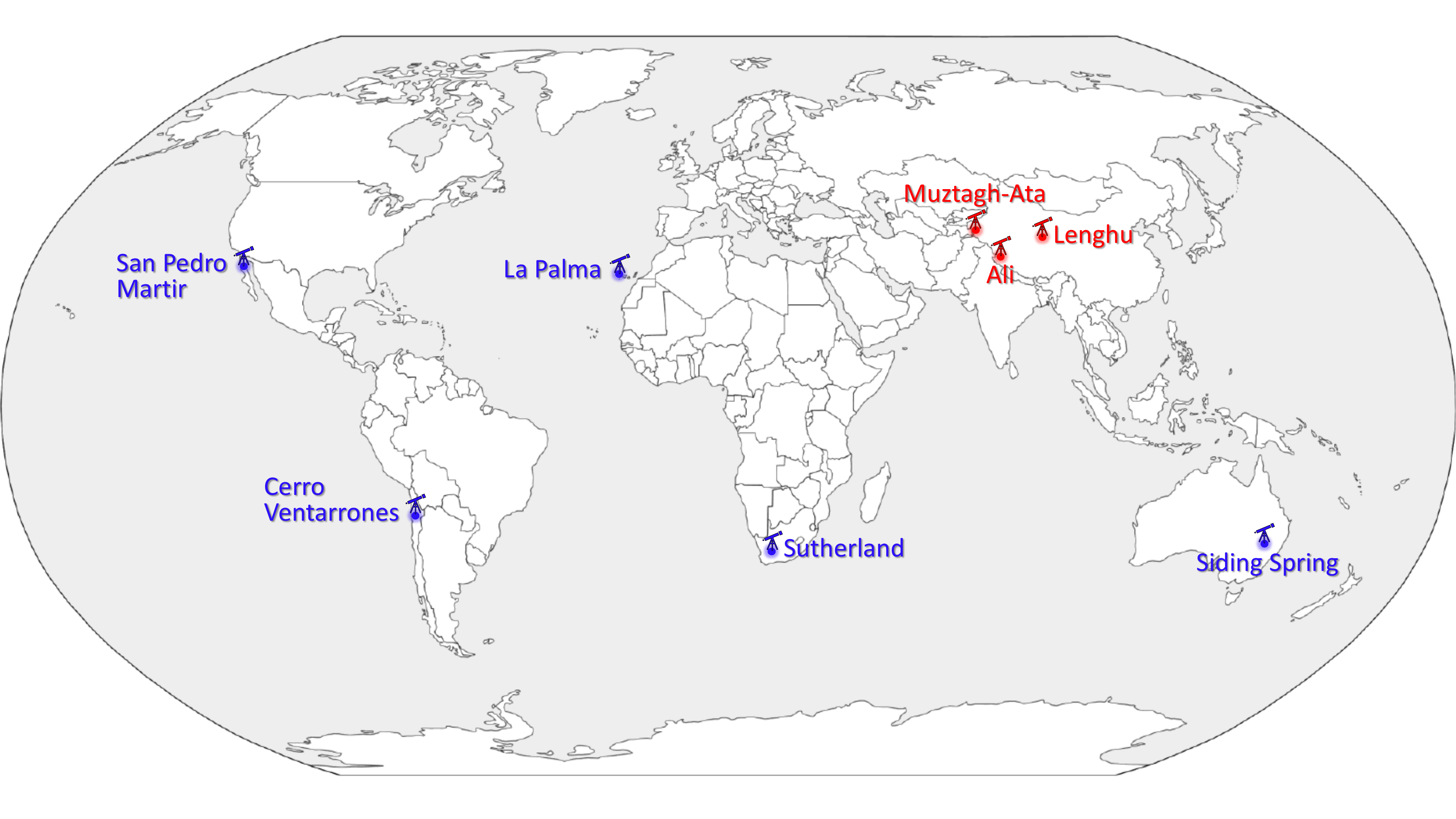}
\caption{Location of the three Chinese sites (in red) confirmed for the first SiTian units, and of candidate sites in other countries (in blue) for the planned subsequent expansion of the network.}
\end{center}
\end{figure}

\vspace{0.3cm}
\noindent
{\large{{\bf{Provisional Timeline}}}}\\
\noindent
We estimate a cost of about US\$300 million for the construction and installation of 72 1-m-class telescopes, not including running costs for operation and maintenance. The specific construction cost of each 1-m telescope is around US\$1.4 million. We budget about US\$100 million for the associated 4-m-class telescopes, and about US\$60 million for the SiTian Brain. We aim to get our project approved and fully funded in the 15th National Mega-Project Five-Year Plan (2026--2030), {\it i.e.}, to start construction from 2026. Before then, the Strategic Planning committee at the National Astronomical Observatories of China (NAOC) has already approved funding for 1 million RMB per year (US\$140,000/yr) for technology development and proof of concept, in preparation for the submission of the SiTian proposal to the 15th Five-Year Plan. If approved, the SiTian project will roughly follow this timeline:\\
2020--2026: overall design layout\\
2026--2029: hardware and software testing\\
2029: completion of infrastructure facilities\\
2030: complete installation of all 72 telescopes\\
2032: full scientific operations\\

\section{Scientific Objectives}\label{objectives}
The key advantage of SiTian over, for example, LSST is its high cadence, which permits the discovery of new transients within 30 min. The main scientific objectives of the project are designed to make use of this feature. However, SiTian will also be useful for studies of other classes of (slower) variable sources, such as stars and exoplanets. Here is a summary of what we hope to find with SiTian.

\vspace{0.3cm}
\noindent
{\bf {1. Electromagnetic counterparts of gravitational waves.}}\\
\noindent
The optical counterpart (``kilonova") of the double neutron star merger event GW170817 was discovered 10 hours after the LIGO detection \cite{abbott17}. The lack of information about the crucial first few hours hampers the modelling of matter ejection and nuclear synthesis processes \cite{pian17}. For future detections of double neutron star (or perhaps neutron star/black hole) mergers, the SiTian array will provide simultaneous three-colour light-curves within half an hour of the gravitational wave event, and will perform rapid follow-up spectroscopic observations. In some cases, it is even possible that the optical detection of a transient at the right time and in the right sector of sky will help confirm a candidate low-signal gravitational wave event. With SiTian for the optical bands, satellites such as {\it Insight-HXMT} \cite{zhang20} and {\it Einstein Probe} \cite{yuan15} for the X-ray bands, and FAST \cite{li13} for the radio bands, in the next decade China will have a broad-band, integrated network for the prompt search of gravitational event counterparts. 

\vspace{0.3cm}
\noindent
{\bf {2. Early light-curves of supernovae (SNe).}}\\
\noindent
The initial shape and colour evolution of the optical light-curves from Type Ia SNe are important factors for the normalization of this type of ``standardizable candles" \cite{riess99}; thus, a better knowledge of the rise phase and colour at peak luminosity will reduce the scatter in the fitted distance, and will put stronger constraints on cosmological parameters. More specifically, very early observations of Type Ia SNe can help distinguish between different types of detonation models and between double-degenerate or single-degenerate progenitors ({\it e.g.}, \cite{blinnikov06,kasen10,li19}).

The {\it Kepler Space Telescope} \cite{haas10} has monitored more than 400 galaxies with a cadence of 30 minutes, and has provided important new information on the early evolution of Type Ia SNe and on their progenitors. However, the Kepler field of view is only 116 square deg, compared with SiTian's 10,000 square deg on the same cadence. Thus, SiTian will dramatically increase the number of observed Type Ia SNe and will have a huge impact on cosmology.

SiTian's early detection will also benefit the study of core-collapse SNe. Those events start to be visible when the shock wave breaks out of the stellar envelope. The gas behind the shock is radiation dominated; the behaviour of the light-curve in the first $\sim$10 hrs depends on whether or not the radiation is in thermal equilibrium. In turn, this depends on the radius and density of the envelope in the massive progenitor star ({\it {e.g.}}, \cite{nakar10,rabinak11,kozyreva20}). SiTian's monitoring of SN light-curves in the first few hours will provide empirical tests for models of shock break-out and will help us constrain the mass range of stars that explode as SNe; thus, indirectly, this will also help us constrain the mass range of stellar black holes.

\vspace{0.3cm}
\noindent
{\bf{3. Intermediate-luminosity optical transients.}}\\
\noindent 
Until a few years ago, there was a clear gap between the optical luminosity ranges of novae and SNe, and it appeared to be relatively simple to identify which events corresponded to stellar explosions and which did not. However, more recently, several examples of optical transients have been found (especially in the near infrared) with intermediate luminosities between novae and SNe. 
For example, ``luminous red novae" are stellar explosions interpreted as the merger of two main sequence stars \cite{kulkarni07}. Electron-capture SNe from low-mass progenitors ($\approx$8--12 solar masses) and the associated process of accretion induced collapse of a Neon-Oxygen white dwarf into a neutron star \cite{dessart06,thompson09,jones16} are another theoretical possibility for optical flares less luminous than classical core-collapse SNe. ``SN impostors" \cite{smith09} are another ill-defined empirical class of events (typically associated with very massive stars near the end of their lives) in this luminosity range; in this case, the optical brightening is due to the ejection of part of the envelope without the complete disruption of the star (for example, giant eruptions of luminous blue variable stars, or systems similar to Eta Carinae). ``Failed SNe" are defined as massive stars in which the core collapsed directly into a black hole without disrupting the envelope; they may produce long-duration optical brightenings and/or a simple disappearance of the star \cite{gerke15}.

SiTian's survey strategy will catch the start of such outbursts within 30 min, and follow them hundreds of times for weeks and months, providing multi-colour lightcurves. As in the case of proper SNe, prompt spectroscopic observations with SiTian's 4-m-class telescopes will help classify the physical type of event.

\vspace{0.3cm}
\noindent
{\bf{4. Tidal disruption events (TDEs).}}\\
\noindent
In general, the initial rise in the light-curve of TDEs constrains the radius at which the star is disrupted and the size of the accretion disk, essential parameters of the star/black holes system ({\it e.g.}, \cite{gezari12,piran15}). Combined optical/UV and X-ray measurements are needed to constrain the total bolometric luminosity. More specifically, there is an unexplained dichotomy between TDE flares discovered in the X-ray band and those discovered in the optical band \cite{hung17}; the former have blackbody temperatures around $10^6$ K, the latter only a few $10^4$ K, with little or no emission in X-rays. It is still unclear whether the two types of TDEs correspond to different ranges of black hole masses, or of Eddington ratios, or different properties of the disrupted stars. Another unsolved issue is the discrepancy between the predicted rate of TDEs and the much lower observed rate \cite{vanvelzen14,stone16}.

The SiTian array is expected to detect about 500 TDEs per year, sampling brightness and colour evolution before the brightness peak, and providing more accurate event rates for different types of galaxies. One of the associated 4-m-class telescopes will be immediately used to obtain spectra, in particular the ratios and widths of H$\alpha$ and He {\footnotesize{II}} lines, which are essential to classify the physical nature of the transient and the speed of the outflows ({\it e.g.}, \cite{arcavi14,roth16}).

\vspace{0.3cm}
\noindent
{\bf{5. Optical counterparts of Gamma-ray bursts (GRBs).}}\\
\noindent
The {\it Neil Gehrels Swift Observatory} \cite{gehrels04} transformed our understanding of GRBs, thanks to the immediate detection and localization of their soft X-ray and optical/UV counterparts; however, several problems remain to be understood, especially for the short GRBs (interpreted as the merger of neutron star binaries). Very few short GRBs have been localized to a specific galaxy, and none within 200 Mpc. Other space missions such as the {\it Fermi} Gamma-ray Burst Monitor \cite{meegan09} have detected many more GRBs (for example, {\it Fermi} detects four times as many GRBs as {\it Swift}) but with large positional uncertainties (tens or even hundreds of square deg); thus, ground-based optical facilities with large fields of view and sufficient cadence and depth are required to identify their counterparts and follow their afterglows over several weeks \cite{graham19,coughlin19}.

The experience of other wide-field monitoring projects such as the ZTF indicates that SiTian will be an excellent facility for such studies. The simultaneous color information provided by SiTian will be especially valuable. A larger sample of optical light-curves and colour evolution will provide empirical tests for models of relativistic jet launching and collimation. It will also enable the development of a unified theoretical model to link the electromagnetic (kilonova) and gravitational-wave signals of neutron star mergers.

\vspace{0.3cm}
\noindent
{\bf{6. Fast radio bursts (FRBs).}}\\
\noindent
FRBs are bright, unresolved radio flashes typically seen in the GHz band, with the duration of up to a few milliseconds \cite{lorimer07,petroff19}. FRBs have been detected so far from about 100 locations in the sky. About 20 of the (unknown) emitters produce recurrent bursts (though only one or two at regular intervals), most of the others do not. Their isotropic distribution in the sky suggests an extragalactic origin. This is also consistent with the high dispersion measure of the bursts (that is, the time delay in the arrivals of different radio frequencies). Their origin is one of the most actively debated mysteries in astronomy today. Many theoretical explanations have been proposed; popular ones include magnetar flares, giant pulses from young pulsars, and neutron star mergers. It is also possible that repeating FRBs have a different origin from non-repeating FRBs, and the two classes may be distinguished by their multiband counterparts. 

No reliable optical counterparts or afterglows have been found for FRBs in external galaxies so far (we are excluding the recently discovered FRB associated with the Galactic magnetar SGR J1935$+$2154: \cite{li20}), and only very few of the bursts have been localized with sufficient accuracy to identify at least a candidate host galaxy \cite{tendulkar17}. Finding the elusive optical counterparts of FRBs is one key goal of any wide-field monitoring survey (for example, see \cite{raiteri18} for such planned search with the LSST). SiTian's higher cadence and larger field of view put it in the best position to discover such counterparts and provide the decisive breakthrough for their physical interpretation.

\vspace{0.3cm}
\noindent
{\bf{7. Ultraluminous X-ray bursts (UXBs).}}\\
\noindent
From an analysis of archival Chandra data of nearby galaxies, \cite{irwin16} found two short point-like flares, which reached peak luminosities of several times $10^{40}$ erg s$^{-1}$ (highly super-Eddington for a stellar-mass black hole) with a rise time of less than a minute. Before and after the flare, those sources appeared as unremarkable low-mass X-ray binaries in old stellar populations. The origin of the flares is still unknown. It is likely that many more UXB flares in nearby galaxies are routinely missed, because they do not reach the hard X-ray threshold for triggering a detection by either the {\it Swift} Burst Alert Telescope or the {\it Fermi} Gamma-ray Burst Monitor. Since such flares have been detected so far only in archival data, there has been no chance to look for their optical counterparts. SiTian will have a chance to constrain the optical brightness of UXBs, especially if its operation overlaps with the lifetime of the {\it Einstein Probe} X-ray telescope.

\vspace{0.3cm}
\noindent
{\bf{8. Variable stars.}}\\
\noindent 
Humble ``normal" stars might seem to provide very little scientific excitement, compared with the mysterious high-energy events described earlier; however, this is far from the truth. As an example of how little we know about normal star variability, we can mention the popular interest in the mysterious dipping behaviour of KIC 8462852 (``Tabby's star": \cite{boyajian16}) or the unexpected dimming of Betelgeuse last year \cite{levesque20}. Learning more about stellar activity cycles from their optical variability may also help us understand the long-term behaviour of the Sun and its effect on the earth's climate.

Even stars with the same temperature and luminosity (that is, at the same position in the HR diagram) may have different optical variability, owing to differences in their internal structure. Only a tiny fraction of Milky Way stars have been regularly monitored for variability, with the same instrument and regular, high cadence. SiTian will monitor millions of stars with 1\% photometric accuracy over many years. It will provide an unprecedented database of variability due to internal cycles, ellipsoidal variations, eclipses, planet transits, flares, etc.

\vspace{0.3cm}
\noindent
{\bf {9. Exoplanets.}}\\
\noindent 
The SiTian array's field of view is 100 times larger than {\it Kepler}'s, and 5 times larger than {\it TESS}'s. A disadvantage of SiTian with respect to {\it Kepler} or {\it TESS} is that the photometric accuracy of SiTian will only reach 1\% for 16-mag stars and 0.1\% for 12-mag stars, compared with better than 1/10,000 for those space telescopes. On the other hand, satellites such as {\it Kepler} and {\it TESS} have a more limited bandwidth and can only transmit to earth a small fraction of the data they collect.
We estimate that SiTian will find about 20,000 exoplanets (five times as much as are currently known) from transit analysis (for stars brighter than 15 mag), and an additional 500 planets from gravitational microlensing of fainter stars. Among all the SiTian findings, we estimate that we will get about 500 Neptune-like and super-earth planets. SiTian is expected to identify about 500 planets around M dwarfs; the size of the habitable zone in that class of stars (a function also of the magnetic activity of the star itself) and the probability of planets in those zones are actively debated topics \cite{vidotto13,airapetian17}.

We want to stress here the suitability of SiTian for {\bf {gravitational microlensing studies}}. The almost continuous monitoring of sources on a fast cadence makes it easier to determine whether any increase in brightness is a one-off event and has a symmetric rise and decline, which are two defining properties of microlensing flares; instead, variable stars usually have periodic oscillations and/or asymmetric flares. Moreover, SiTian's strategy of simultaneous observations in three colours makes it easier to determine whether the events are achromatic (the third defining property of microlensing, because lensing is independent of wavelength). In addition to a new sample of short-period exoplanets, the microlensing search will enable SiTian to discover short-period white dwarf - white dwarf and white dwarf - neutron star binaries, as well as stray gas planets, brown dwarfs and isolated compact objects passing across our line of sight to a more distant star.

\vspace{0.3cm}
\noindent
{\bf{10. Solar system objects.}}\\
\noindent 
SiTian's scanning mode (10,000 deg field, 21 mag limit, 30 min cadence) is very suitable for the search, monitoring, and precise orbit determination of potentially hazardous asteroids and near-earth objects \cite{harris15,jones18}, including also the brightest end of the distribution of temporarily captured objects (popularly known as Earth's transient moons; \cite{fedorets20}). Simultaneous imaging in three colours will facilitate the classification of at least some of those near earth objects into stony, carbonaceous and metallic types, removing uncertainties due to rotation; this has promising applications to asteroid prospecting and mining \cite{elvis17,graps19}. More detailed physical and chemical properties of such objects will be obtained from follow-up spectroscopic studies with the associated 4-m-class telescopes. SiTian will also obtain light-curves of main-belt asteroids, from which we can infer their shape, rotation period, and other information. In addition, SiTian is also expected to discover and monitor thousands of Kuiper belt objects \cite{petit11}.

\section{Benefits to society}\label{outreach}
A long-term project such as SiTian benefits other areas of society, outside pure research. Cataloging and monitoring of near-earth objects and space debris are two direct applications of interest to national security and to commercial enterprises. SiTian's true-colour ``movie" of a variable sky will be an excellent tool for science outreach, and will produce huge social benefits in terms of general scientific education; astronomy discoveries and images often motivate talented young people to pursue a career in STEM. In addition, the SiTian Brain will require the development of new computing and data management resources and services, which will have applications also in other fields with more direct applications to society ({\it e.g.}, weather and climate forecasting, land surveying and management, epidemiology). The challenge of collecting, storing, processing, selecting and distributing huge amounts of data and information within a few minutes will be an opportunity to collaborate with high-tech industries. The international nature of the project (including student exchanges and staff visits) will have a positive impact on diplomatic relations and on the reputation of Chinese science overseas.

As a specific example of how SiTian is already having an outreach dimension, we would like to mention the involvement of the Eighth High School in Beijing (and soon of other high schools) in the current phase of software and hardware testing. School students are currently writing software for the automatic operation of a 50-cm telescope, with simultaneous three-colour imaging (via dichroics). The field of view is $0^\circ.6 \times 0^\circ.6$ for each of the three cameras, and is designed to match GSENSE4040 detectors. (GSENSE404 is a 16.8 Megapixel CMOS detector, with 9$\mu$m pixels, readout noise of 3.7$e^{-}$ in rolling shutter mode. It is produced by the Gpixel company, with headquarters in Changchun, Northern China.) Once in full operation, SiTian will actively involve high school students for educational and research projects.

\section{Conclusions}\label{conclusions}
SiTian's main steps forward will be its 30-min scan cadence over at least 10,000 square deg of sky (5-$\sigma$ detection limit of 21 mag), its simultaneous imaging in three colours, and its access to 4-m telescopes specifically allocated for follow-up spectroscopic studies. In addition to its primary goal of transient discovery and monitoring, SiTian will provide stacked data over various periods of time (for example, reaching 26 mag for a 1-year stack), and will enable research on distant galaxies, faint galaxy halos, AGN, quasars, intracluster stars, etc.

Currently, the SiTian project (PI: Prof.~JiFeng Liu) has nearly one hundred team members from 24 Chinese institutes, including NAOC (Chinese Academy of Sciences), Tsinghua University and Peking University in Beijing, and Purple Mountain Observatory in Nanjing. Thus, it is a project with a high support and participation within the Chinese astronomy community. We are currently starting to build a support network of institutions and observatories outside China who may be interested in joining the SiTian team. Project membership will guarantee full and immediate access to the raw and pipeline calibrated data and to in-house computational and processing resources.

SiTian's design has comparatively low technical risks. The technology for the 1-m-class and 4-m-class telescopes is well tested and relatively mature in China. Fully robotic Chinese telescopes are already in operation or construction in Antarctica (Antarctic Survey Telescopes). sCMOS detector technology is developing fast, with falling prices and improving performances, partly driven by mobile phone camera requirements. The amount of data generated by SiTian will reach about 100 PB per year, which is comparable to the data rate of LSST. It requires careful design, planning and testing of hardware and software tools for data collection, processing, transmission, storage and fast access; however, it is not an insurmountable problem. We will build on the local experience of large data management for other astronomical surveys in China, such as the Beijing-Arizona Sky Survey (wide-field multi-colour survey of 5000 square deg around the northern Galactic pole), and of large surveys overseas, such as the Square Kilometer Array and its associated Pawsey Supercomputing Centre, in Perth.  Technical and financial support for data management from commercial giants such as Alibaba and Tencent will also be available. Finally, our project will gain invaluable experience from international collaborations and from our participation in other wide-field surveys over the next decade. To name just a few: the Gattini and DREAMS infrared surveys at Palomar and Siding Spring, respectively \cite{moore19,de20}; the PRime-focus Infrared Microlensing Experiment (PRIME) from the 1.8-m telescope at SAAO; the SAAO's African Intelligent Observatory network; the future BRICS Intelligent Telescope and Data Network.

\section{Acknowledgments}\label{acknowledgments}
We thank the members of the SiTian team, and in particular Dr Wang Song, for technical support in the preparation of this summary report. We also thank Prof.~David Buckley for comments, technical discussions and warm hospitality at SAAO, and for his visits to NAOC in Beijing. We are grateful to the organizers of the 2019 BRICS Astronomy workshop in Rio for a successful and productive meeting. We appreciated the careful reading of this manuscript by the anonymous referee. RS thanks the University of Sydney for their hospitality and support during part of this work. This report has greatly benefited from our interaction and discussions with Paul Groot, Mansi Kasliwal, Shri Kulkarni, Meng Meng, Stephen Potter, Shuang-Nan Zhang, Tianmeng Zhang.   

\bibliographystyle{abbrv}
\bibliography{main}

\begin{thebibliography}{10}

\bibitem{abbott17}
B.~P. {Abbott} et~al.
\newblock {Multi-messenger Observations of a Binary Neutron Star Merger}.
\newblock {\em \apjl}, 848(2):L12, Oct. 2017.

\bibitem{airapetian17}
V.~S. {Airapetian}, A.~{Glocer}, G.~V. {Khazanov}, R.~O.~P. {Loyd},
  K.~{France}, J.~{Sojka}, W.~C. {Danchi}, and M.~W. {Liemohn}.
\newblock {How Hospitable Are Space Weather Affected Habitable Zones? The Role
  of Ion Escape}.
\newblock {\em \apjl}, 836(1):L3, Feb. 2017.

\bibitem{arcavi14}
I.~{Arcavi} et~al.
\newblock {A Continuum of H- to He-rich Tidal Disruption Candidates With a
  Preference for E+A Galaxies}.
\newblock {\em \apj}, 793(1):38, Sept. 2014.

\bibitem{bellm19}
E.~C. {Bellm} et~al.
\newblock {The Zwicky Transient Facility: System Overview, Performance, and
  First Results}.
\newblock {\em \pasp}, 131(995):018002, Jan. 2019.

\bibitem{blinnikov06}
S.~I. {Blinnikov}, F.~K. {R{\"o}pke}, E.~I. {Sorokina}, M.~{Gieseler},
  M.~{Reinecke}, C.~{Travaglio}, W.~{Hillebrand t}, and M.~{Stritzinger}.
\newblock {Theoretical light curves for deflagration models of type Ia
  supernova}.
\newblock {\em \aap}, 453(1):229--240, July 2006.

\bibitem{boyajian16}
T.~S. {Boyajian} et~al.
\newblock {Planet Hunters IX. KIC 8462852 - where's the flux?}
\newblock {\em \mnras}, 457(4):3988--4004, Apr. 2016.

\bibitem{brown13}
T.~M. {Brown} et~al.
\newblock {Las Cumbres Observatory Global Telescope Network}.
\newblock {\em \pasp}, 125(931):1031, Sept. 2013.

\bibitem{coughlin19}
M.~W. {Coughlin} et~al.
\newblock {2900 Square Degree Search for the Optical Counterpart of Short
  Gamma-Ray Burst GRB 180523B with the Zwicky Transient Facility}.
\newblock {\em \pasp}, 131(998):048001, Apr. 2019.

\bibitem{de20}
K.~{De} et~al.
\newblock {Palomar Gattini-IR: Survey Overview, Data Processing System, On-sky
  Performance and First Results}.
\newblock {\em \pasp}, 132(1008):025001, Feb. 2020.

\bibitem{dessart06}
L.~{Dessart}, A.~{Burrows}, C.~D. {Ott}, E.~{Livne}, S.~C. {Yoon}, and
  N.~{Langer}.
\newblock {Multidimensional Simulations of the Accretion-induced Collapse of
  White Dwarfs to Neutron Stars}.
\newblock {\em \apj}, 644(2):1063--1084, June 2006.

\bibitem{elvis17}
M.~{Elvis}.
\newblock {Astronomical Prospecting of Asteroid Resources}.
\newblock In {\em European Planetary Science Congress}, pages EPSC2017--94,
  Sept. 2017.

\bibitem{fedorets20}
G.~{Fedorets}, M.~{Granvik}, R.~L. {Jones}, M.~{Juri{\'c}}, and R.~{Jedicke}.
\newblock {Discovering Earth's transient moons with the Large Synoptic Survey
  Telescope}.
\newblock {\em \icarus}, 338:113517, Mar. 2020.

\bibitem{gehrels04}
N.~{Gehrels} et~al.
\newblock {The Swift Gamma-Ray Burst Mission}.
\newblock {\em \apj}, 611(2):1005--1020, Aug. 2004.

\bibitem{gerke15}
J.~R. {Gerke}, C.~S. {Kochanek}, and K.~Z. {Stanek}.
\newblock {The search for failed supernovae with the Large Binocular Telescope:
  first candidates}.
\newblock {\em \mnras}, 450(3):3289--3305, July 2015.

\bibitem{gezari12}
S.~{Gezari} et~al.
\newblock {An ultraviolet-optical flare from the tidal disruption of a
  helium-rich stellar core}.
\newblock {\em \nat}, 485(7397):217--220, May 2012.

\bibitem{graham19}
M.~J. {Graham} et~al.
\newblock {The Zwicky Transient Facility: Science Objectives}.
\newblock {\em \pasp}, 131(1001):078001, July 2019.

\bibitem{graps19}
A.~L. {Graps} et~al.
\newblock {ASIME 2018 White Paper. In-Space Utilisation of Asteroids: Asteroid
  Composition -- Answers to Questions from the Asteroid Miners}.
\newblock {\em arXiv e-prints}, page arXiv:1904.11831, Apr. 2019.

\bibitem{haas10}
M.~R. {Haas} et~al.
\newblock {Kepler Science Operations}.
\newblock {\em \apjl}, 713(2):L115--L119, Apr. 2010.

\bibitem{harris15}
A.~W. {Harris} and G.~{D'Abramo}.
\newblock {The population of near-Earth asteroids}.
\newblock {\em \icarus}, 257:302--312, Sept. 2015.

\bibitem{hung17}
T.~{Hung} et~al.
\newblock {Revisiting Optical Tidal Disruption Events with iPTF16axa}.
\newblock {\em \apj}, 842(1):29, June 2017.

\bibitem{irwin16}
J.~A. {Irwin} et~al.
\newblock {Ultraluminous X-ray bursts in two ultracompact companions to nearby
  elliptical galaxies}.
\newblock {\em \nat}, 538(7625):356--358, Oct. 2016.

\bibitem{ivezic19}
{\v{Z}}.~{Ivezi{\'c}} et~al.
\newblock {LSST: From Science Drivers to Reference Design and Anticipated Data
  Products}.
\newblock {\em \apj}, 873(2):111, Mar. 2019.

\bibitem{jones18}
R.~L. {Jones} et~al.
\newblock {The Large Synoptic Survey Telescope as a Near-Earth Object discovery
  machine}.
\newblock {\em \icarus}, 303:181--202, Mar. 2018.

\bibitem{jones16}
S.~{Jones}, F.~K. {R{\"o}pke}, R.~{Pakmor}, I.~R. {Seitenzahl}, S.~T.
  {Ohlmann}, and P.~V.~F. {Edelmann}.
\newblock {Do electron-capture supernovae make neutron stars?. First
  multidimensional hydrodynamic simulations of the oxygen deflagration}.
\newblock {\em \aap}, 593:A72, Sept. 2016.

\bibitem{kasen10}
D.~{Kasen}.
\newblock {Seeing the Collision of a Supernova with Its Companion Star}.
\newblock {\em \apj}, 708(2):1025--1031, Jan. 2010.

\bibitem{keller07}
S.~C. {Keller} et~al.
\newblock {The SkyMapper Telescope and The Southern Sky Survey}.
\newblock {\em \pasa}, 24(1):1--12, May 2007.

\bibitem{kochanek17}
C.~S. {Kochanek} et~al.
\newblock {The All-Sky Automated Survey for Supernovae (ASAS-SN) Light Curve
  Server v1.0}.
\newblock {\em \pasp}, 129(980):104502, Oct. 2017.

\bibitem{kozyreva20}
A.~{Kozyreva}, E.~{Nakar}, R.~{Waldman}, S.~{Blinnikov}, and P.~{Baklanov}.
\newblock {Shock breakouts from red supergiants: analytical and numerical
  predictions}.
\newblock {\em \mnras}, 494(3):3927--3936, Apr. 2020.

\bibitem{kulkarni07}
S.~R. {Kulkarni} et~al.
\newblock {An unusually brilliant transient in the galaxy M85}.
\newblock {\em \nat}, 447(7143):458--460, May 2007.

\bibitem{levesque20}
E.~M. {Levesque} and P.~{Massey}.
\newblock {Betelgeuse Just Is Not That Cool: Effective Temperature Alone Cannot
  Explain the Recent Dimming of Betelgeuse}.
\newblock {\em \apjl}, 891(2):L37, Mar. 2020.

\bibitem{li20}
C.~K. {Li} et~al.
\newblock {Identification of a non-thermal X-ray burst with the Galactic
  magnetar SGR 1935+2154 and a fast radio burst with Insight-HXMT}.
\newblock {\em arXiv e-prints}, page arXiv:2005.11071, May 2020.

\bibitem{li13}
D.~{Li}, R.~{Nan}, and Z.~{Pan}.
\newblock {The Five-hundred-meter Aperture Spherical radio Telescope project
  and its early science opportunities}.
\newblock In J.~{van Leeuwen}, editor, {\em Neutron Stars and Pulsars:
  Challenges and Opportunities after 80 years}, volume 291 of {\em IAU
  Symposium}, pages 325--330, Mar. 2013.

\bibitem{li19}
W.~{Li} et~al.
\newblock {Photometric and Spectroscopic Properties of Type Ia Supernova 2018oh
  with Early Excess Emission from the Kepler 2 Observations}.
\newblock {\em \apj}, 870(1):12, Jan. 2019.

\bibitem{lorimer07}
D.~R. {Lorimer}, M.~{Bailes}, M.~A. {McLaughlin}, D.~J. {Narkevic}, and
  F.~{Crawford}.
\newblock {A Bright Millisecond Radio Burst of Extragalactic Origin}.
\newblock {\em Science}, 318(5851):777, Nov. 2007.

\bibitem{lou16}
Z.~{Lou} et~al.
\newblock {Optical design study of the Wide Field Survey Telescope (WFST)}.
\newblock In {\em \procspie}, volume 10154 of {\em Society of Photo-Optical
  Instrumentation Engineers (SPIE) Conference Series}, page 101542A, Oct. 2016.

\bibitem{meegan09}
C.~{Meegan} et~al.
\newblock {The Fermi Gamma-ray Burst Monitor}.
\newblock {\em \apj}, 702(1):791--804, Sept. 2009.

\bibitem{moore19}
A.~M. {Moore} and M.~M. {Kasliwal}.
\newblock {Unveiling the dynamic infrared sky}.
\newblock {\em Nature Astronomy}, 3:109--109, Jan. 2019.

\bibitem{nakar10}
E.~{Nakar} and R.~{Sari}.
\newblock {Early Supernovae Light Curves Following the Shock Breakout}.
\newblock {\em \apj}, 725(1):904--921, Dec. 2010.

\bibitem{petit11}
J.~M. {Petit} et~al.
\newblock {The Canada-France Ecliptic Plane Survey{\textemdash}Full Data
  Release: The Orbital Structure of the Kuiper Belt}.
\newblock {\em \aj}, 142(4):131, Oct. 2011.

\bibitem{petroff19}
E.~{Petroff}, J.~W.~T. {Hessels}, and D.~R. {Lorimer}.
\newblock {Fast radio bursts}.
\newblock {\em \aapr}, 27(1):4, May 2019.

\bibitem{pian17}
E.~{Pian} et~al.
\newblock {Spectroscopic identification of r-process nucleosynthesis in a
  double neutron-star merger}.
\newblock {\em \nat}, 551(7678):67--70, Nov. 2017.

\bibitem{piran15}
T.~{Piran}, G.~{Svirski}, J.~{Krolik}, R.~M. {Cheng}, and H.~{Shiokawa}.
\newblock {Disk Formation Versus Disk Accretion{\textemdash}What Powers Tidal
  Disruption Events?}
\newblock {\em \apj}, 806(2):164, June 2015.

\bibitem{rabinak11}
I.~{Rabinak} and E.~{Waxman}.
\newblock {The Early UV/Optical Emission from Core-collapse Supernovae}.
\newblock {\em \apj}, 728(1):63, Feb. 2011.

\bibitem{raiteri18}
C.~M. {Raiteri} et~al.
\newblock {Blazars and Fast Radio Bursts with LSST}.
\newblock {\em arXiv e-prints}, page arXiv:1812.03151, Dec. 2018.

\bibitem{riess99}
A.~G. {Riess} et~al.
\newblock {The Rise Time of Nearby Type IA Supernovae}.
\newblock {\em \aj}, 118(6):2675--2688, Dec. 1999.

\bibitem{roth16}
N.~{Roth}, D.~{Kasen}, J.~{Guillochon}, and E.~{Ramirez-Ruiz}.
\newblock {The X-Ray through Optical Fluxes and Line Strengths of Tidal
  Disruption Events}.
\newblock {\em \apj}, 827(1):3, Aug. 2016.

\bibitem{smith09}
N.~{Smith} et~al.
\newblock {SN 2008S: A Cool Super-Eddington Wind in a Supernova Impostor}.
\newblock {\em \apjl}, 697(1):L49--L53, May 2009.

\bibitem{soon20}
J.~{Soon} et~al.
\newblock {Wide-field dynamic astronomy in the near-infrared with Palomar
  Gattini-IR and DREAMS}.
\newblock In {\em \procspie}, volume 11203 of {\em Society of Photo-Optical
  Instrumentation Engineers (SPIE) Conference Series}, page 1120307, Jan. 2020.

\bibitem{stone16}
N.~C. {Stone} and B.~D. {Metzger}.
\newblock {Rates of stellar tidal disruption as probes of the supermassive
  black hole mass function}.
\newblock {\em \mnras}, 455(1):859--883, Jan. 2016.

\bibitem{tendulkar17}
S.~P. {Tendulkar} et~al.
\newblock {The Host Galaxy and Redshift of the Repeating Fast Radio Burst FRB
  121102}.
\newblock {\em \apjl}, 834(2):L7, Jan. 2017.

\bibitem{thompson09}
T.~A. {Thompson}, J.~L. {Prieto}, K.~Z. {Stanek}, M.~D. {Kistler}, J.~F.
  {Beacom}, and C.~S. {Kochanek}.
\newblock {A New Class of Luminous Transients and a First Census of their
  Massive Stellar Progenitors}.
\newblock {\em \apj}, 705(2):1364--1384, Nov. 2009.

\bibitem{vanvelzen14}
S.~{van Velzen} and G.~R. {Farrar}.
\newblock {Measurement of the Rate of Stellar Tidal Disruption Flares}.
\newblock {\em \apj}, 792(1):53, Sept. 2014.

\bibitem{vidotto13}
A.~A. {Vidotto}, M.~{Jardine}, J.~{Morin}, J.~F. {Donati}, P.~{Lang}, and
  A.~J.~B. {Russell}.
\newblock {Effects of M dwarf magnetic fields on potentially habitable
  planets}.
\newblock {\em \aap}, 557:A67, Sept. 2013.

\bibitem{yuan15}
W.~{Yuan} et~al.
\newblock {Einstein Probe - a small mission to monitor and explore the dynamic
  X-ray Universe}.
\newblock {\em arXiv e-prints}, page arXiv:1506.07735, June 2015.

\bibitem{zhang20}
S.-N. {Zhang} et~al.
\newblock {Overview to the Hard X-ray Modulation Telescope (Insight-HXMT)
  Satellite}.
\newblock {\em Science China Physics, Mechanics, and Astronomy}, 63(4):249502,
  Feb. 2020.

\end{thebibliography}

\end{document}